\documentclass[11pt,tightenlines,amsmath,amssymb]{revtex4}
\usepackage{graphicx,dcolumn,bm}% Include figure files
\newlength{\minitwocolumn}

\def\linebreak{\hfill\break}

\def\bra<#1|{\langle #1\rvert}
\def\ket|#1>{\lvert#1 \rangle}
\def\braket<#1|#2>{\langle #1|#2 \rangle}

\def\pfrac#1#2{\left(\frac{#1}{#2}\right)}

\def\tend{\rightarrow}

\def\therefore{\mbox{\setbox0=\hbox{X}\hbox{$\ldotp$}\raise0.7\ht0\hbox{$\ldotp$}\hbox{$\ldotp$}} \quad }
\def\because{\mbox{\setbox0=\hbox{X}\raise0.7\ht0\hbox{$\ldotp$}\hbox{$\ldotp$}\raise0.7\ht0\hbox{$\ldotp$}}\kern0pt }

\def\r#1{{\rm #1}}

\def\Frac(#1/#2){\left(\frac{#1}{#2}\right)}

\def\Order#1{\r{O}\!\left(#1\right)}

\def\Eq#1{\begin{equation} #1 \end{equation}}

\def\Eqr#1{\begin{eqnarray} #1 \end{eqnarray}}

\def\Eqrsub#1{\begin{subequations}
\Eqr{#1}\end{subequations}}
\def\Eqrsubl#1#2{\begin{subequations}\label{#1}
\Eqr{#2}\end{subequations}}
\def\Bitm{\begin{itemize}}
\def\Eitm{\end{itemize}}
\def\Blist#1#2{\begin{list}{#1}{\parsep=0pt \itemsep=0pt%   
  \listparindent=0pt #2}}
\def\Elist{\end{list}}

\begin{document}
\title{Global structure of the Zipoy-Voorhees-Weyl spacetime\\
and the $\delta=2$ Tomimatsu-Sato spacetime}
\author{Hideo Kodama} 
\email{kodama@yukawa.kyoto-u.ac.jp}
\author{Wataru Hikida}
\email{hikida@yukawa.kyoto-u.ac.jp}

\affiliation{Yukawa Institute for Theoretical Physics, Kyoto University, Kyoto 606-8502, Japan}

%T1>Abstract 
\begin{abstract}
We investigate the structure of the ZVW (Zipoy-Voorhees-Weyl) 
spacetime, which is a Weyl solution described by the Zipoy-Voorhees 
metric, and the $\delta=2$ Tomimatsu-Sato spacetime. We show that 
the singularity of the ZVW spacetime, which is represented by a 
segment $\rho=0, -\sigma<z<\sigma$ in the Weyl coordinates, is geometrically 
point-like for $\delta<0$, string-like for $0<\delta<1$ and 
ring-like for $\delta>1$. These singularities are always naked and 
have positive Komar masses for $\delta>0$. Thus, they provide a 
non-trivial example of naked singularities with positive mass. We 
further show that the ZVW spacetime has a degenerate Killing horizon 
with a ring singularity at the equatorial plane for $\delta=2,3$ and 
$\delta\ge4$. We also show that the $\delta=2$ Tomimatsu-Sato 
spacetime has a degenerate horizon with two components, in contrast 
to the general belief that the Tomimatsu-Sato solutions with even 
$\delta$ do not have horizons.
\end{abstract}
\preprint{}
\maketitle

%T1>Introduction
%%=========================================================== 
\section{Introduction}
%%-------------------------------------------

Almost thirty years have passed since Tomimatsu and 
Sato\cite{Tomimatsu.A&Sato1972} found a series of exact solutions, 
the so-called TS solutions, to the vacuum Ernst equation, which can 
be regarded as representing quadrupole deformations of the Kerr 
solution. However, the global structure of these solutions has not 
been studied well, or at least, its details are not known widely, 
though some basic features such as the existence of naked ring 
singularities and causality violating regions are well 
known\cite{Tomimatsu.A&Sato1973,Tomimatsu.A&Sato1973a,%
Gibbons.G&Russell-Clark1973}. Furthermore, some conflicting or wrong 
statements on their global and singularity structures are sometimes 
found in the literature for the case the deformation parameter 
$\delta$ is even. For example, Gibbons and 
Russell-Clark\cite{Gibbons.G&Russell-Clark1973} claimed that the TS 
(Tomimatsu-Sato) solution with $\delta=2$, which is referred to as 
TS2 in the present paper, does not have a horizon, by showing that 
the spacetime is inextendible across the segment I: 
$\rho=0,-\sigma<z<\sigma$ in the Weyl coordinates, which corresponds 
to a horizon for odd 
$\delta$\cite{Tomimatsu.A&Sato1973a}. Although they noticed that TS2 
has quasi-regular directional singularities at P$_\pm$: 
$(\rho,z)=(0,\pm\sigma)$ in the Weyl coordinates, and that the 
spacetime has two-dimensional extensions along the $z$-axis across 
these points, they failed to recognize that these points are 
horizons. Subsequently, Ernst\cite{Ernst.F1976} pointed out that 
these points are hypersurfaces, and investigated the behavior of 
geodesics crossing them by introducing a polar coordinate system 
around these points. However, he gave the incorrect conclusion that 
the corresponding surfaces are time-like and a particle which 
crossed one of the hypersurfaces can come out again through the same 
hypersurface. Later, Papadopoulos et 
al\cite{Papadopoulos.D&Stewart&Witten1981} recognized that the two 
points P$_\pm$ of the ZVW spacetime with $\delta\ge2$, which is a Weyl 
solution described by the Zipoy-Voorhees metric\cite{Zipoy.D1966,Voorhees.B1970,Stephani.H&&2003B} and  a static 
limit of the TS solution for integer values of $\delta$, are also 
hypersurfaces and tried to construct a coordinate system providing a 
smooth extension of the spacetime across these hypersurfaces. They 
found a two-dimensional extension along the $z$-axis, but were not 
able to find a coordinate system giving a four-dimensional 
extension. 

One main reason for the lack of detailed studies may be the fact 
that the TS solutions and the Weyl solutions have pathological 
features including the existence of naked 
singularities. If the cosmic censorship 
hypothesis\cite{Penrose.R1969} were correct, such solutions would 
not describe the real world. However, the cosmic censorship is still 
a hypothesis, and there are now lots of counter examples%
\cite{Papapetrou.A1985B,Hollier.G1986,Singh.T1996A,Gundlach.C2000A,%
Harada.T&Iguchi&Nakao2002}. Of course, it is probable that the 
cosmic censorship hypothesis turns out to be correct in practice, 
since most of the analytic counter examples found so far are 
spherically symmetric%
\cite{Wald.R1997A}. Hence, in order to clarify whether the cosmic 
censorship is correct or not, it is inevitable to study 
non-spherically symmetric systems\cite{Lake.K1992}. For that 
purpose, knowledge on possible structures of spacetimes with naked 
singularities will give useful information. Such knowledge will be 
also important if the cosmic censorship turns out to be incorrect. 
From this point of view, in the present paper, we investigate in 
detail the global structure of the ZVW and TS spacetimes, with focus 
on singularities and horizons. Although we do not restrict the 
deformation parameter $\delta$ for the ZVW solution, we only 
consider the $\delta=2$ case for the TS solution, because it is hard 
to analyse the TS solutions with $\delta\ge3$, even by using 
symbolic computations by computers. 

We will reveal lots of new features of the spacetimes described by 
these solutions concerning the singularity and causal structure. In 
particular, we will prove that the ZVW spacetimes with $\delta=2,3$ 
and $\delta\ge4$ have a degenerate Killing horizon at P$_\pm$, as 
anticipated by Papadopoulos et al, and show that the horizon has a 
ring singularity at the equator. We further show that at least for 
$\delta=2$, the Tomimatsu-Sato spacetime has degenerate Killing 
horizons at P$_\pm$, each of which is a sphere with conic 
singularity. We will also point out that the ZVW solution with 
$\delta>0$ ($\delta\not=1$) provides a rare example of non-trivial 
naked singularity with positive gravitational mass.

The paper is organized as follows. In section II, we first study in 
detail the structure of the singularities of the ZVW spacetimes, and 
then show that for $\delta=2,3$ and $\delta\ge4$, the 
directional quasi-regular singularities at P$_\pm$ are regular parts 
of a degenerate Killing horizon, by giving explicit regular 
extensions of the spacetime across these points. In section III, we 
extend the analysis to TS2. After a brief review of the basic 
features of this spacetime, we prove that the 
directional quasi-regular singularities at P$_\pm$ are degenerate 
Killing horizons. We also clarify the structure of singularity at 
the segment I on the $z$-axis. Section IV is devoted to summary and 
discussion. All symbolic computations in the present paper were done 
by Maple V.

%T1>Zipoy-Voorhees-Weyl spacetime
%%-------------------------------------------
\section{Zipoy-Voorhees-Weyl Spacetime}
%%-------------------------------------------

The ZVW spacetimes are a two-parameter family of static, 
axisymmetric and asymptotically flat vacuum solutions to the 
Einstein equations described by the Zipoy-Voorhees 
metric\cite{Zipoy.D1966,Voorhees.B1970,Stephani.H&&2003B}, which 
include the flat and Schwarzschild solutions. In the canonical form 
for the static axisymmetric metric
\Eq{
 ds^2=-fdt^2+f^{-1}[e^{2\gamma}(d\rho^2+dz^2)+\rho^2d\phi^2],
\label{ZVW:metric:(rho,z)}
}
they are expressed as\cite{Voorhees.B1970}
\Eq{
 f=\left(\frac{x-1}{x+1}\right)^{\delta}, \quad 
 e^{2\gamma}=\left(\frac{x^2-1}{x^2-y^2}\right)^{\delta^2}.
}
Here, $x$ and $y$ are the prolate spheroidal coordinates related to the Weyl coordinates $(\rho, z)$ by 
\Eq{
\rho=\sigma\sqrt{(x^2-1)(1-y^2)},\quad z=\sigma xy. \label{eq:prolate}
}
In this prolate spheroidal coordinate system $(t,\phi,x,y)$, the 
ZVWmetric is written as
\Eq{
ds^2=-fdt^2+R^2d\phi^2
      +\Sigma^2\left(\frac{dx^2}{x^2-1}+\frac{dy^2}{1-y^2}\right),
\label{ZVW:metric:(x,y)}
}
where
\Eqr{
&& R^2=\sigma^2 \left(\frac{x+1}{x-1}\right)^{\delta-1}(x+1)^2(1-y^2),
\label{ZVW:R:xy}\\
&& \Sigma^2=\sigma^2\frac{(x+1)^{\delta^2+\delta}}
        {(x-1)^{-\delta^2+\delta}}(x^2-y^2)^{1-\delta^2}.
}

Here, note that the metric is invariant under the transformation 
$\delta \rightarrow -\delta, x\rightarrow -x$. Further, the metric 
has a curvature singularity at some value of $x$, except for the 
case $\delta=0$, as we see below. Hence, we can assume that the 
asymptotically flat region corresponds to the range $x>1$ and $-1\le 
y\le 1$. In this asymptotic flat region, for the spherical 
coordinates $(r,\theta,\phi)$ defined by $\sigma x=r-\delta\sigma 
+{\rm O}(1/r)$ and $y=\cos\theta$,  the metric components in 
\eqref{ZVW:metric:(x,y)} are expanded as 
\Eqrsubl{ZVW:SpatialInfinity}{
&& f=1 -\frac{2\sigma \delta}{r} + \Order{\frac{1}{r^3}},\\
&& R^2/\sin^2\theta=r^2+2\sigma \delta r
   + (4\delta^2-1)\sigma^2+\Order{\frac{1}{r}},\\
&& \Sigma^2=r^2+2\sigma\delta r + [4\delta^2-1+(1-\delta^2)\sin^2\theta]\sigma^2 
   +\Order{\frac{1}{r}}.
}
From these expressions, we see that the gravitational mass of the 
system  is represented as $M=\delta \sigma$, and that the parameter 
$\delta$ taking a real number describes a quadrupole deformation of 
the spacetime\cite{Zipoy.D1966,Voorhees.B1970}: each $f=$constant 
surface has a prolate shape for $|\delta|<1$ and an oblate shape for 
$|\delta|>1$. In particular, $\delta=0$ corresponds to the flat 
metric, and $\delta=\pm1$ to the Schwarzschild metric with positive 
or negative mass.

\setlength{\minitwocolumn}{0.45\textwidth}
\begin{figure}[t]
%T3>fig:Weylstructure
\begin{minipage}{\minitwocolumn}
  \begin{center}
  \includegraphics*[height=\minitwocolumn]{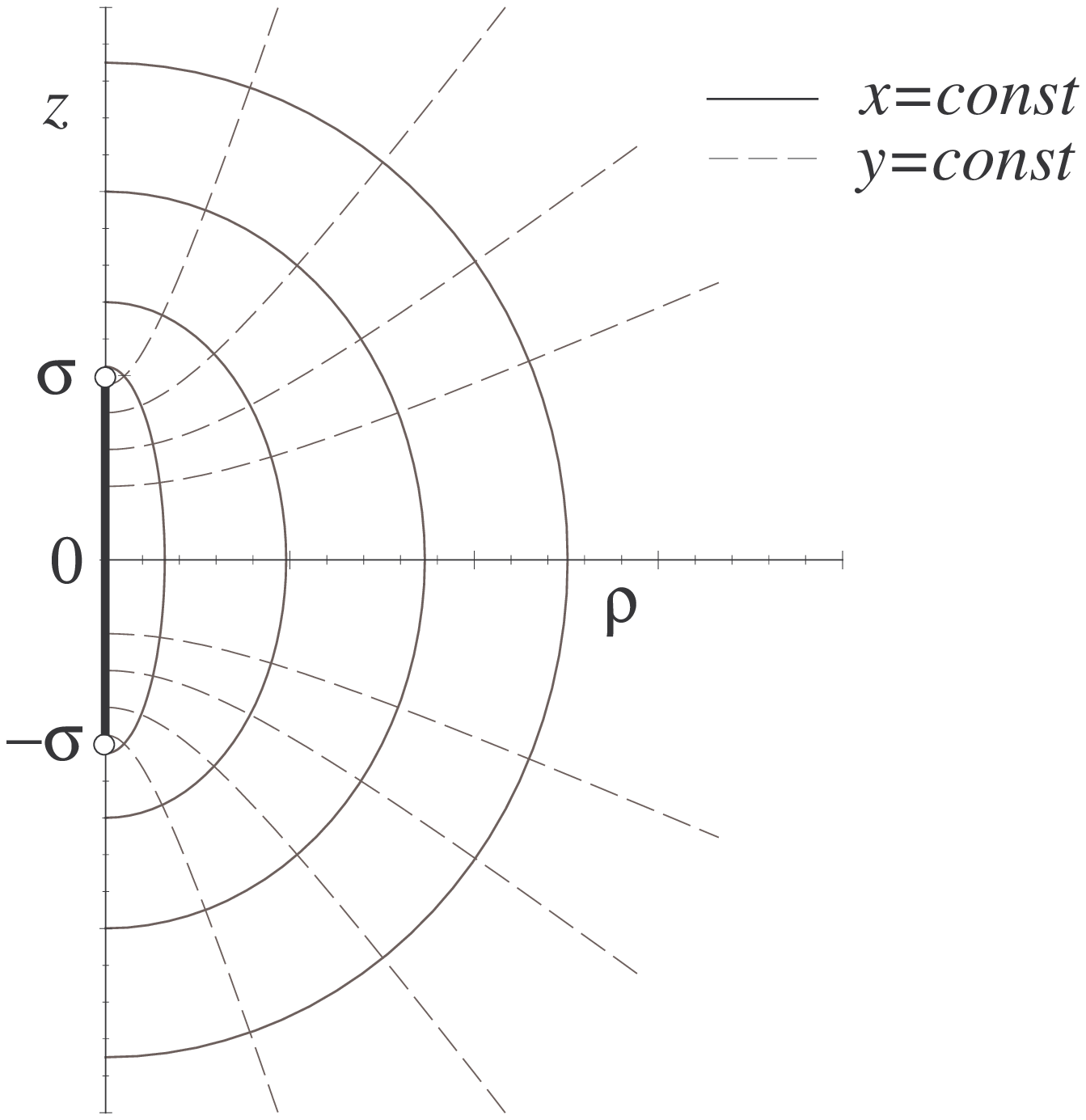}
  \caption{The structure of the ZVW spacetime. The thick solid 
	 line and the two open circles denote the curvature singularity 
	 and the directional singularities, respectively. The $x=$const and 	 	 $y=$const curves are also shown. }
	 \label{fig:Weylstructure}
  \end{center}
\end{minipage}
\hspace{0.05\textwidth}
%T3>fig:XYcoord
\begin{minipage}{\minitwocolumn}
  \begin{center}
  \includegraphics*[height=\minitwocolumn]{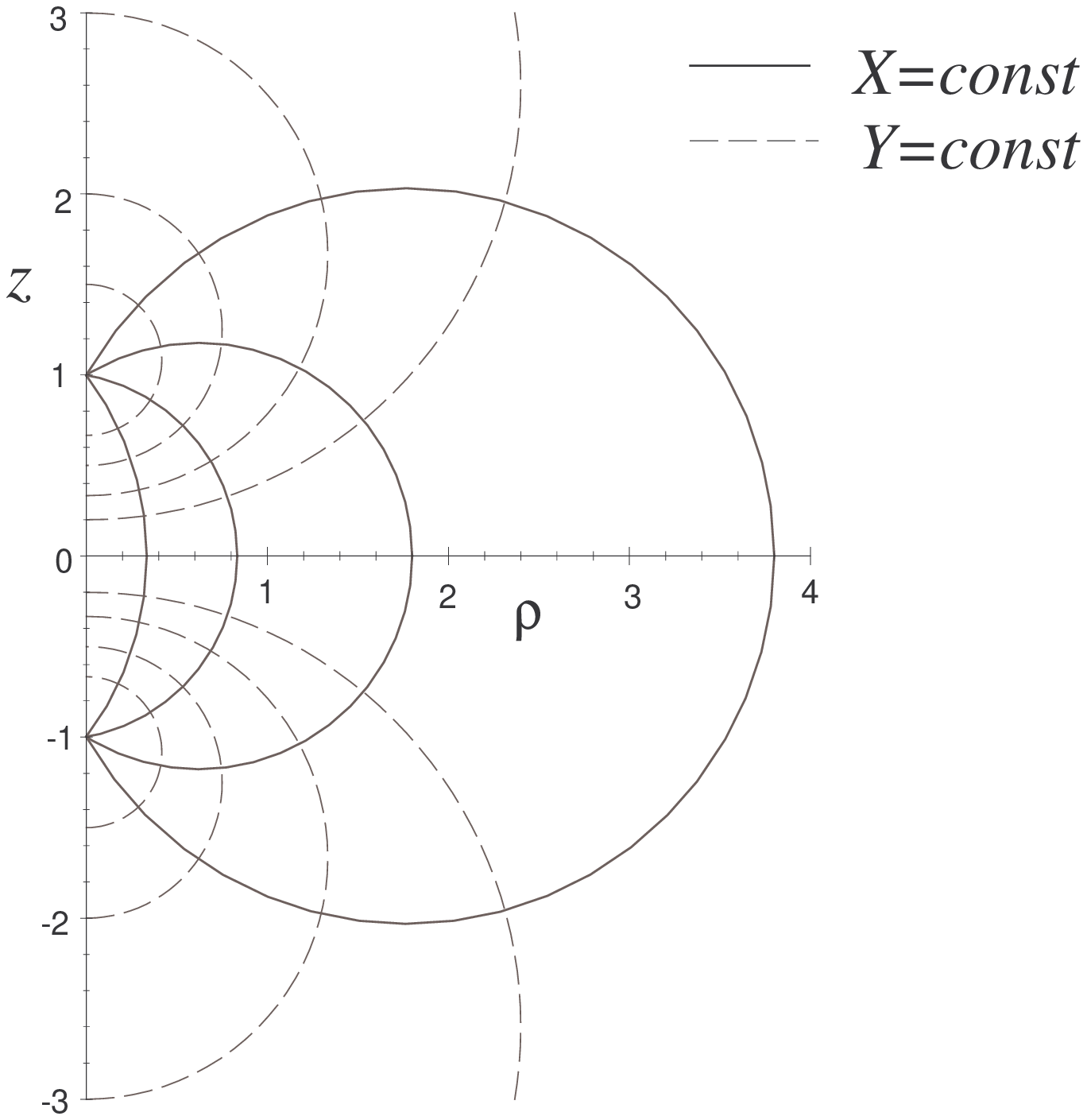} 
  \caption{The relation between the $(\rho,z)$ and the $(X,Y)$ 
       coordinate systems in the units $\sigma=1$. The solid and the 
       dotted curves are $X=$const and $Y=$const curves, 
       respectively. }
	 \label{fig:XYcoord}
      \end{center}
\end{minipage}
\end{figure}

%T2>Singularity structure
\subsection{Singularity structure}

%T3>Curvature invariants
From \eqref{ZVW:metric:(x,y)}, it is clear that the metric can have 
curvature singularity only at $x=\pm1$ and $x=\pm y$. In this 
section, we first confirm that there is actually singularity at 
$x=\pm1$ for $\delta\not=0,1$ and investigate the structure of the 
singularity there. In the present paper, we focus on singularities 
accessible from the asymptotically flat region $x>1, -1\le y\le1$, 
and do not consider possible singularities at $x=\pm y$, which are 
separated from the asymptotically flat region by naked singularities 
or horizons.

For a static and axisymmetric metric, $\Psi_1=\Psi_3=0$ and all 
scalar polynomials of the Weyl curvature can be written as 
polynomials of $\Psi_2$ and $\Psi_0\Psi_4$, where $\Psi_0\sim\Psi_4$ 
are the Newman-Penrose components of the Weyl curvature with respect 
to the complex null tetrad $(k,l,m,\bar{m})$ defined by% 
\Eq{
    k_\mu dx^\mu = \frac{\sqrt{f}dt+Rd\phi}{\sqrt{2}},\quad
   l_\mu dx^\mu = \frac{\sqrt{f}dt-Rd\phi}{\sqrt{2}},\quad
   m_\mu dx^\mu = \frac{e^\gamma(d\rho-idz)}{\sqrt{2f}}.
\label{ZVW:NullTetrad}
}
For the metric \eqref{ZVW:metric:(x,y)}, these curvature invariants are given by
\Eqrsubl{ZVW:curvature:invariants}{
&& \Psi_2=\frac{\delta (x-\delta)(x^2-y^2)^{\delta^2-1}}
          {2(x-1)^{\delta^2-\delta+1}(x+1)^{\delta^2+\delta+1}},\\
&\Psi_0\Psi_4
=& \frac{\delta^2(x^2-y^2)^{2\delta^2-3}}
    {4[(x-1)^{\delta^2-\delta+1}(x+1)^{\delta^2+\delta+1}]^2}
    \times\Big[9(x-1)^2(x^2-y^2)
 \nonumber\\
&&     
+(\delta-1)\big\{[12(\delta+1)x^2-6(2\delta^2+2\delta+3)x+(\delta+1)(4
\delta^2+5)](1-y^2)
 \nonumber\\
&& -9(2x-\delta-1)(x^2-1)\big\}\Big].
}
%

\begin{figure}[t]
\begin{minipage}{7cm}
\includegraphics*[height=6cm]{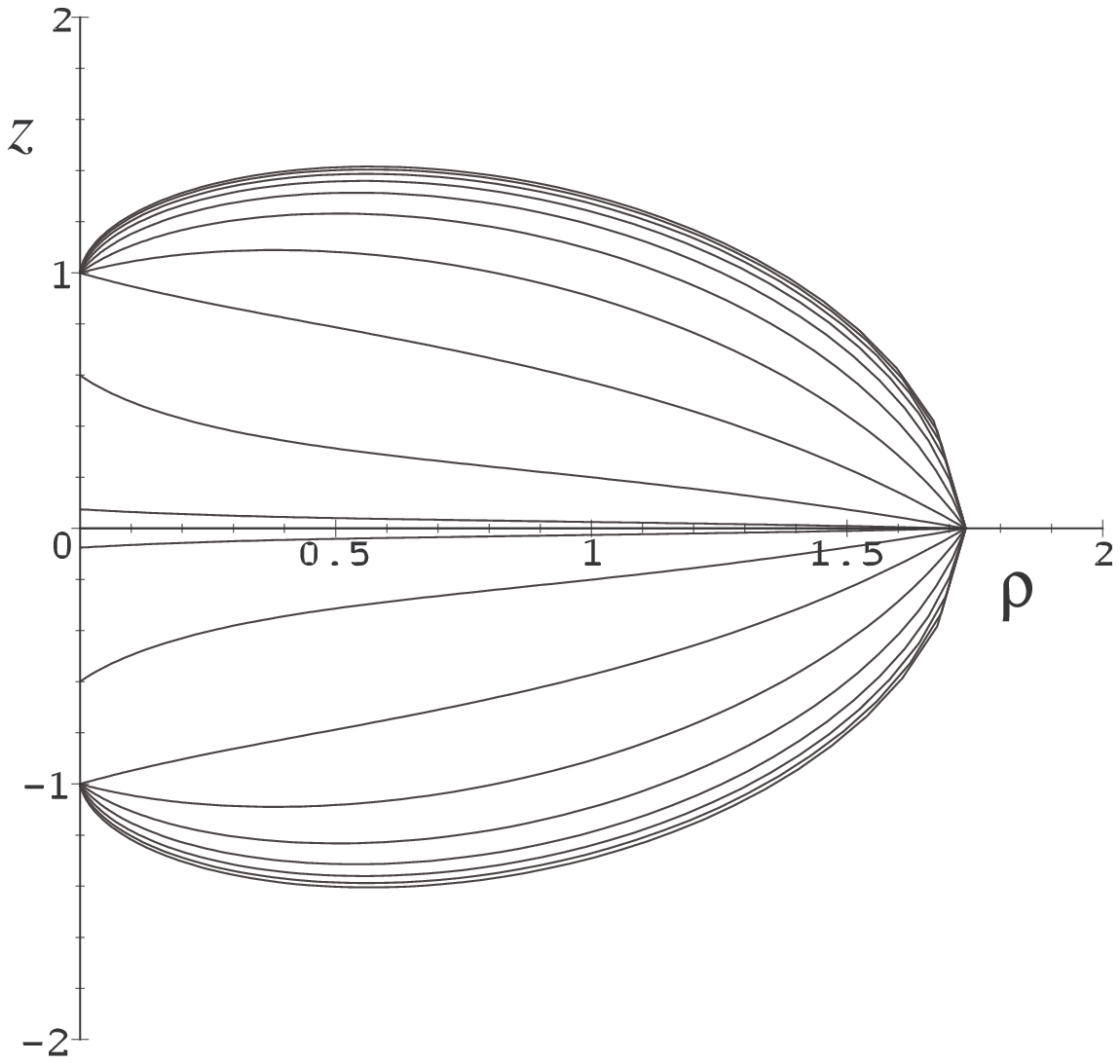}
\end{minipage}
\begin{minipage}{7cm}
\includegraphics*[height=6cm]{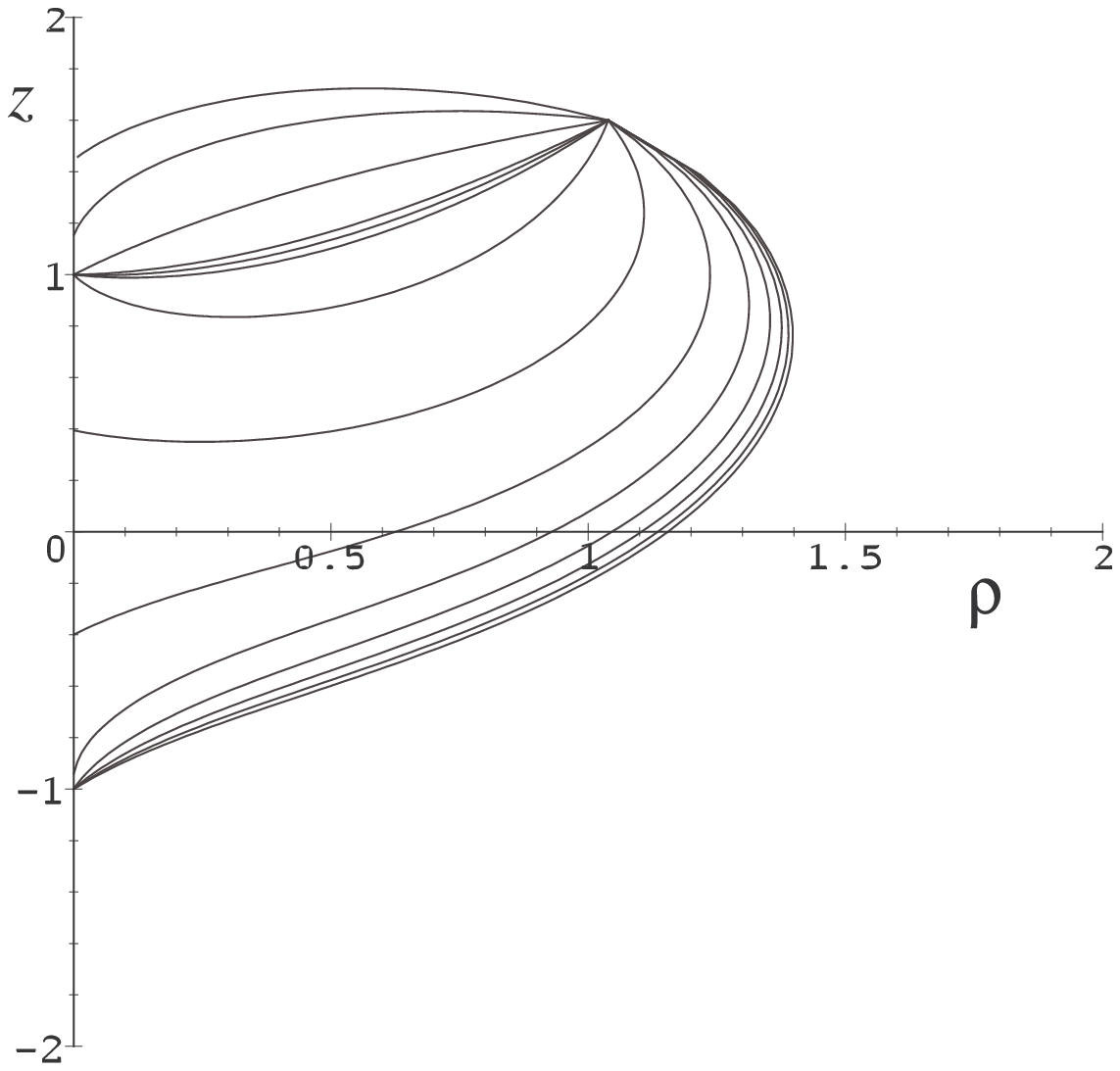}
\end{minipage}
\caption{\label{fig:Geodesics:delta=2}Two examples of a family of 
null geodesics reaching the singular segment I from the same point 
for $\delta=2$. These figures are drawn in the units $\sigma=1$.}
\end{figure}

%T3>Segment I
From these expressions, we immediately see that the curvature 
invariants diverge on the open segment I: $x=1, -1<y<1$, if 
$\delta\not=0,1$. This segment corresponds to the segment $\rho=0,
 -\sigma<z<\sigma$ in the Weyl coordinates, which is depicted by a 
thick line in Fig.\ref{fig:Weylstructure}. It is also easy to see 
that one can reach any point on this segment by a curve with a 
finite proper length in the $x-y$ plane. As is expected from this, 
this segment is a naked singularity. In fact, we can show that in 
the case $L=0$ and $\epsilon=0$ (null geodesics), the geodesic 
equation \eqref{ZVW:GeodesicEq:orbit} has a regular solution 
expressed by a power series of $x-1$ as 
\Eq{
y=y_0+\frac{\delta+1}{\delta-1}y_0(x-1)+\cdots,
}
for any $y_0$ in the open interval $(-1,1)$, provided that 
$\delta\not=1$. From \eqref{ZVW:GeodesicEq:affine}, we see that the 
null geodesic represented by this solution reaches the point 
$(x,y)=(1,y_0)$ in a finite affine parameter. This implies that each 
point in the open segment $\rho=0, -\sigma<z<\sigma$ is a naked 
curvature singularity in the standard sense. By looking for a 
solution of the type $y=y_0+a(x-1)^p+\cdots$ ($p>0$) to 
\eqref{ZVW:GeodesicEq:orbit}, we can also show that for $\delta<0$ 
or $\delta>2$ there exists no other solution that reaches the point 
$(1,y_0)$ for $y_0^2<1$, while for $0<\delta<1$ or $1<\delta\le2$ 
there exist one parameter family of such solutions with 
$p=(1+2\delta-\delta^2)/2$. Hence, for the latter case, one can 
reach every point on the open segment I from a point with $x>1$ by a 
null geodesic. Fig. \ref{fig:Geodesics:delta=2} gives two examples 
of such a family of null geodesics for the case $\delta=2$.  

%T3>fig:Geodesics:delta=2

%T3>Points P+-
The curvature invariants also exhibit singular behavior at the two 
points P$_\pm$: $(\rho,z)=(0,\pm\sigma)$. To see this, we introduce 
the new coordinate system $(X,Y)$ defined by 
\Eq{
X^2=\frac{1-y^2}{x^2-1},\quad
Y=\frac{y}{x}.
\label{XYcoord:def}}
The coordinates $x$ and $y$ are expressed in terms of $X$ and $Y$ as
\Eq{
x=\sqrt{\frac{X^2+1}{X^2+Y^2}},\quad
y=Y\sqrt{\frac{X^2+1}{X^2+Y^2}}.
}
The region $x\ge1, |y|\le1$ is mapped to the region $X\ge0, 
|Y|\le1$, and $X=$const curves and $Y=$const curves are represented 
by circles with centers on the $\rho$-axis and on the $z$-axis, 
respectively, as shown in Fig.\ref{fig:XYcoord}.  

Inserting these expressions into \eqref{ZVW:curvature:invariants}, 
we find that the curvature invariants behave in the limit $Y^2\tend 
1$ with fixed $X$ as 
\Eqrsub{
&& \Psi_2 \approx \frac{\delta(1-\delta)}{2^{2\delta+1}}
    \left(1+X^2\right)^{\delta^2-\delta+1}(1-Y^2)^{\delta-2},\\
&& \Psi_0\Psi_4 \approx \frac{\delta^2(\delta-1)^2}{2^{4\delta+4}}
    \left(1+X^2\right)^{2\delta^2-2\delta+1}
    \left(9+(2\delta-1)^2X^2\right)
    \left(1-Y^2\right)^{2(\delta-2)}.
}
From this, we see that the two points P$_\pm$(i.e., $Y=\pm1$) are 
curvature singularities for $\delta<2$ ($\delta\not=0,1$). In this 
case, one can reach each of these points by a spatial curve as well 
as by a null geodesic along the $z$-axis in the $\rho-z$ plane with 
a finite affine length $\ell$, because it follows from 
\eqref{ZVW:GeodesicEq:affine} that the affine parameter is 
proportional to the $x$-coordinate for such geodesics. Hence, they 
are naked.  

In contrast, for $\delta>2$, the curvatures vanish as one 
approaches P$_\pm$. Hence, they are not curvature singularity. 
Furthermore, there exists no spatial curve that reaches these points 
within a finite proper length, as was pointed out by Papadopoulos et 
al\cite{Papadopoulos.D&Stewart&Witten1981}. However, since null 
geodesics along the $z$-axis reach these points in finite affine 
parameters, they are quasi-regular singularity for the region 
$x>1$\cite{Papadopoulos.D&Stewart&Witten1981}. In the next 
subsection, we will show that these points are in reality Killing 
horizons at least for $\delta=3$ and $\delta\ge4$.

The situation for $\delta=2$ is  similar but more subtle. The 
curvature invariants approach non-vanishing values in the limit 
$Y\tend\pm1$, and these limits depend on $X$, i.e., on the direction 
one approaches the points P$_\pm$. They are again at spatial 
infinity, but can be reached by null geodesics with finite affine 
parameters. Hence, they are apparently directional quasi-regular 
singularities\cite{Papadopoulos.D&Stewart&Witten1981}. We will also 
show that these points are Killing horizons.

%T3>tbl:WeylSingularity
\begin{table}
\begin{tabular}{c|c|c|c|c|c|c|c|c|c|c}
\hline
$\delta$ 
 &\multicolumn{2}{|c|}{Curvature} 
 & \multicolumn{2}{|c|}{Radius} 
 &\multicolumn{3}{|c|}{Length}
 &\multicolumn{2}{|c|}{Proper distance}
 & Affine distance
 \\
\cline{2-11}
    & I & P$_\pm$
    & I & P$_\pm$ 
    & $L_x$ & $L_X$ & $L_n$
    & I & P$_\pm$ 
    & I/P$_\pm$
    \\
\hline
$\delta<0$ 
  & $\infty$ & $\infty$
  & 0 & 0 
  & 0 & 0 & 0 
  & finite & finite 
  & finite
  \\
$\delta=0$ 
  & 0 & 0 
  & 0 & 0 
  & $2\sigma$ & $2\sigma$& $2\sigma$ 
  & finte & finite
  & finite
  \\
$0<\delta<1$
  & $\infty$ & $\infty$ 
  & 0 & 0 
  & $\infty$ & $\infty$ & $\infty$ 
  & finite & finite
  & finite
  \\
$\delta=1$ 
  & finite & finite
  & 0 & finite 
  & $2\pi\sigma$ & $2\pi\sigma$ & $2\pi\sigma$ 
  & finite & finite
  & finite
  \\
$1<\delta<2$ 
  & $\infty$ & $\infty$
  & $\infty$ & 0 
  & 0 & 0 & 0 
  & finite & finite 
  & finite
  \\
$\delta=2$ 
  & $\infty$ & finite
  & $\infty$ & $4\sigma X$ 
  & $8\sigma$ & $\infty$ & 0 
  & finite & $\infty$ 
  & finite
  \\
$\delta>2$
  & $\infty$ & 0
  & $\infty$ & $\infty$ 
  & $\infty$ & $\infty$ & 0
  & finite & $\infty$ 
  & finite
  \\
\hline
\end{tabular}
\caption{\label{tbl:WeylSingularity} Asymptotic behavior of the ZVW 
spacetime near the singularity. The second and third columns show 
the limits of the curvature invariants on the open segment I: 
$\rho=0,-\sigma<z<\sigma$ and at the two points P$_\pm$: 
$(\rho,z)=(0,\pm\sigma)$. The next two columns with the heading 
'Radius' show the limit of $R$ on the open segment I and at the two 
points P$_\pm$, respectively, and the following three columns with 
the heading 'Length' show the limit at $x=1$ of the length of the 
curves in the $x-y$ plane with $x=$const, $X=$const and 
$X(1-Y^2)^n=$const, respectively. Finally, the two columns denoted 
by 'Proper distance'  show the proper length $\ell$ of spatial 
curves that approach a point on I and the two points P$_\pm$, 
respectively, and the last column shows the behaviour of the  affine 
length of null geodesics approaching the singularities.} \end{table}

%T3>Geometrical shape
Next, let us examine geometrical shapes of these singularities. One 
simple quantity characterizing shapes is the circumferential radius 
$R$ of circles generated by the Killing vector $\eta=\partial_\phi$, 
given by \eqref{ZVW:R:xy}.  The asymptotic behavior of this 
quantity near the singularity is summarized in Table 
\ref{tbl:WeylSingularity}. This asymptotic behavior suggests that 
the singularity is point-like or string like for $\delta<1$ 
($\delta\not=0$) and is ring-like for $\delta>1$. 

To confirm this, we estimate the length of the singularity in the 
$z$-direction. There are various candidates for the definition of 
such a length. One natural candidate is the length of $x=$const 
curves,%
\Eqr{
&L_x &:=\int_{-1}^1\frac{\Sigma dy}{\sqrt{1-y^2}}\notag\\
&&=\sigma (x+1)^{\delta(\delta+1)/2}
    (x-1)^{\delta(\delta-1)/2}
    \int_0^{\pi/2}d\theta(x^2-1+\sin^2\theta)^{(1-\delta^2)/2}.
}
The second candidate is the length of $X=$const curves, which is 
expressed in the limit $X\tend +\infty$ as 
\Eq{
L_X \approx 2^\delta \sigma X^{\delta(1-\delta)}
             \int_{-1}^1\frac{dY}{(1-Y^2)^{\delta/2}}.
}
Here, the integral on the right-hand side diverges for $\delta\ge2$, 
hence $L_X$ can have a finite limit only for $\delta<2$. The 
asymptotic behavior of these quantities in the limits $x\tend1$ and 
$X\tend+\infty$ are summarized in Table \ref{tbl:WeylSingularity}. 
Comparison of this behavior of $L_x$ and $L_X$ with that of $R$ 
shows that the singularity is point-like for $\delta<0$, string-like 
for $0<\delta<1$, and ring-like for $1<\delta<2$. In contrast, these 
quantities do not seem to describe the length of singularities for 
$\delta\ge2$, because the two points P$_\pm$ are horizons and not 
singularity, as we will show later. For these cases, we have to 
measure the length of the singularity by the limit length of curves 
which are tangent to the segment I at $(\rho,z)=(0,\pm\sigma)$. 
Therefore, we define  the length of I by the $a\tend\infty$ limit of 
the length $L_n$ of curves with $X(1-Y^2)^n=a$(constant) for $n>0$. 
In this limit, $L_n$ is expressed as
\Eq{
L_n \approx 2^\delta\sigma a^{\delta(1-\delta)}
     \int_{-1}^1 dY\, (1-Y^2)^{n\delta(\delta-1)-\delta/2}.
}
The integral on the right-hand side of this equation converges for 
$2-\delta-2n\delta(1-\delta)>0$. For any $\delta$, there exists 
$n>0$ that satisfies this inequality, and for such $n$, the 
asymptotic behavior of $L_n$ is simply determined by the sign of 
$\delta(1-\delta)$, as summarized in Table 
\ref{tbl:WeylSingularity}. This behavior implies that the 
singularity is ring-like for $\delta\ge2$.  

To summarize, the segment singularity I is geometrically point-like 
for $\delta<0$, string-like for $0<\delta<1$ and ring-like for 
$\delta>1$. This feature is consistent with the behavior 
\eqref{ZVW:SpatialInfinity} of the metric at spatial infinity for 
$\delta>0$, although such a good correspondence does not exist for 
$\delta<0$. 

Here, note that this characterization of the singularity is based on 
the purely local geometry and may not coincide with its physical 
appearance. For example, if an observer in the region $x>1$ looks at 
the singularity by light rays, he will observe a cylinder-like 
object for the case $1<\delta\le2$, because light rays starting from 
different points on I can reach the same point in $x>1$. For the 
other values of $\delta$, the geometrical shape and the physical 
appearance coincide. 

%T3>Komar mass
Finally, we examine the Komar mass of the singularity. The Komar 
mass is in general defined in terms of the time-like Killing vector 
$\xi=\partial_t$ as\cite{Komar.A1959,Komar.A1962,Heusler.M1996B}
\Eq{
M=-\frac{1}{16\pi}\int_\Sigma \epsilon_{\mu\nu\lambda\sigma}
 \nabla^\mu\xi^\nu dx^\lambda\wedge dx^\sigma.
\label{KomarMass:def}
}
Here, the orientation of the spatial two-surface $\Sigma$ should be 
chosen so that for a future-directed time-like vector $u$Can 
outward normal $n$ to $\Sigma$ and a basis $(v,w)$ giving the 
orientation of $\Sigma$,  $(u,n,v,w)$ gives the positive orientation 
of the spacetime.

For the ZVW metric, in the case in which $\Sigma$ is an axisymmetric 
surface determined by a curve $\gamma$ in the $x-y$ plane, this 
expression reduces to%
\Eq{
M=\frac{\delta\sigma}{2}\int_\gamma dy,
}
where the orientation of $\gamma$ should be taken so that for the 
tangent vector $V$ of $\gamma$ and a normal $N$ to $\gamma$ directed 
outward from the region enclosed by $\gamma$, $(N,V)$ has the same 
orientation as that of $(\partial_x,\partial_y)$. It is clear from 
this expression that each portion of the segment singularity has a 
positive mass and the total mass carried by the singularity 
coincides with the total mass of the system, $M=\delta\sigma$. This 
result is quite interesting because most of the known examples of 
naked singularities have negative or zero 
mass\cite{Singh.T1996A,Gergely.L2002}. For example, central 
shell-focusing singularities formed by spherically symmetric 
gravitational collapse always have 
non-positive local gravitational mass if they are locally 
naked\cite{Lake.K1992}. One exception is a  shell-crossing 
singularity, which can have a positive mass. However, this type of 
singularity is weak in the sense that the spacetime has a $C^0$ 
extension across the singularity\cite{Newman.R1986}. Another counter 
example is the time-like singularity of the Janis-Newman-Winicour 
solution\cite{Janis.A&Newman&Winicour1968,Virbhadra.K1997} for the 
spherically symmetric Einstein-Massless-Scalar (EMS) system. 
Virbhadra showed that this singularity has a positive mass if one 
defines the mass by the spatial integration of the Einstein 
energy-momentum complex\cite{Virbhadra.K1999}. This mass is 
identical to the Komar mass, whose value is independent of the 
location of the integration surface $\Sigma$ for a static solution 
of the spherically symmetric EMS system. However, it does not 
coincide with the local gravitational mass used in the mass 
inflation argument\cite{Poisson.E&Israel1990}. We can show that the 
latter becomes negatively infinite for the singularity of the 
Janis-Newman-Winicour solution. This discrepancy seems to come from 
the difference in the matter contribution to the mass. In contrast, 
because the ZVW solution discussed in the present paper is a vacuum 
solution, this kind of ambiguity associated with the matter 
contribution does not exist in the definition of the mass for the 
singularity of the ZVW spacetime. Thus, it seems to provide a 
clearer non-trivial example of naked singularities with positive 
gravitational mass. 

%T2>Extension across P+-
\subsection{Extension across P$_\pm$}

%T3>fig:SWcoord
\begin{figure}[t]
\includegraphics*[height=8cm]{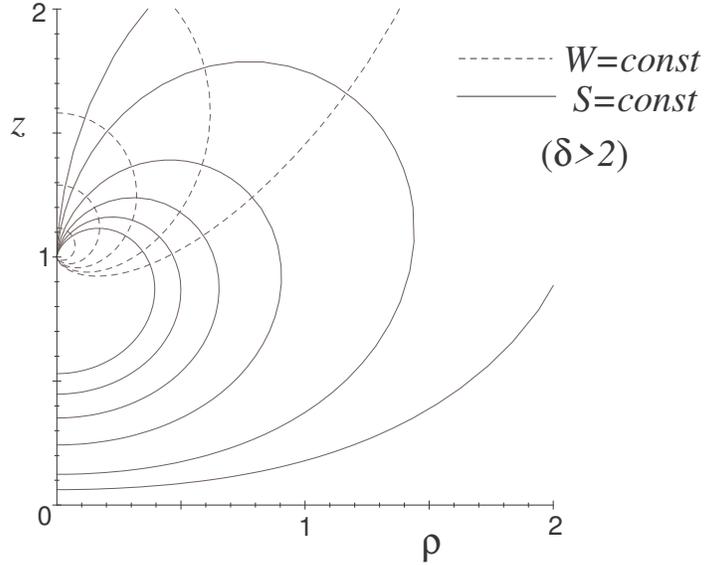}
\caption{\label{fig:SWcoord} The relation between the 
$(S,W)$-coordinates and $(\rho,z)$-coordinates for $\delta>2$. The 
coordinate values of $\rho$ and $z$ are given in the units 
$\sigma=1$.}\end{figure}

 In this subsection,  we will demonstrate that the two points 
 P$_\pm$ are Killing horizons for $\delta=2,3$ and $\delta\ge4$, as 
 anticipated by Papadopoulos et al\cite{Papadopoulos.D&Stewart&Witten1981}, by constructing smooth 
 extensions of the spacetime across these points.

The coordinate system around the points P$_\pm$ ($(X,Y)=(0,\pm1)$) that we adopt to extend the spacetime is 
\Eqrsub{
&& S=2^\delta\sigma \frac{X}{\sqrt{1+X^2}}
     \pfrac{1-Y^2}{Y^2}^{(2-\delta)/2},\\
&& W=\frac{1}{4}(1-Y^2)(1+X^2)^{\delta-2}.
}
For $\delta=2$, $S$ and $W$ depend only on $X$ and $Y$, 
respectively, and these two  coordinate systems are essentially 
equivalent. However, for $\delta>2$, each $S=$const curve is 
tangential to the $z$-axis in the $\rho-z$ plane, as is shown in 
Fig. \ref{fig:SWcoord}. Hence, the $(S,W)$ coordinates expand cusp 
regions at P$_\pm$ in the $(X,Y)$ coordinates. Such a singular 
coordinate system is required because $R$ diverges as $Y$ tends 
$\pm1$ along  $X=$const curves for $\delta>2$, as is shown in Table 
\ref{tbl:WeylSingularity}. 

In this new coordinate system, the ZVW metric is expressed as%
\Eq{
ds^2=-a W^\delta\left( dt^2-h^2\frac{dW^2}{W^{2\delta}}\right)
     +bS^2d\phi^2+cdS^2,
}
with
\Eqrsub{
&& a=\pfrac{2}{x+1}^{2\delta}(X^2+Y^2)^{-\delta}
     (1+X^2)^{\delta(2-\delta)},\\
&& b=\pfrac{x+1}{2}^{2\delta}Y^{2(2-\delta)}(1+X^2)
     (X^2+Y^2)^{\delta-2},\\
&& c=\pfrac{x+1}{2}^{2\delta}\frac{(1+X^2)^{3-\delta^2}
      (X^2+Y^2)^{\delta-2}}{Y^{2(\delta-3)}
      [Y^2+(\delta-2)^2X^2]},\\
&& h^2=4\sigma^2\pfrac{x+1}{2}^{4\delta}
     \frac{(1+X^2)^{\delta^2-6\delta+6}
       (X^2+Y^2)^{2(\delta-1)}}{Y^2+(\delta-2)^2X^2}.
}
Here, the dependence of $x$, $X$ and $Y$ on $S$ and $W$ is implicitly  determined by the relations

\Eqrsub{
&& x=\sqrt{\frac{X^2+1}{X^2+Y^2}},\\
&& Y^2=1-\frac{4W}{(1+X^2)^{\delta-2}},\\
&& X^2=\frac{S^2}{16\sigma^2}\frac{(1+X^2)W^{\delta-2}}
        {[(1+X^2)^{\delta-2}-4W]^{\delta-2}}.
\label{xYXbySW}
}

This coordinate system covers only a neighbourhood of P$_+$ or P$_-$ 
in general, and $P_\pm$ corresponds to $W=0$.  For $\delta=2$,  
$a,b$ and $c$ have non-vanishing finite limits at $W=0$, which are 
expressed as regular functions of $X$, or equivalently of $S$. On 
the other hand, for $\delta>2$, in the limit $W\tend0$ with $S$ 
fixed, $X^2$ behaves as 
\Eq{
X^2=\frac{S^2}{16\sigma^2}W^{\delta-2}\left(1+\Order{W}\right).
}
Hence, $a,b$ and $c$ approach unity in the same limit. 

Here, note that if $\delta$ is an integer equal to or greater than 
2, $a,b$ and $c$ have unique regular analytic extensions to the 
region $W<0$. A $C^2$ extension is possible even for non-integer 
$\delta$ by replacing $W^\delta$ in \eqref{xYXbySW} by $|W|^\delta$, 
if $\delta\ge4$.  

In the extension of the spacetime, a crucial point is the asymptotic 
behavior of $h$ at $W=0$. First, we examine it for $\delta>2$. Let 
us define the function $h_0(W)$ by 
\Eq{
h_0(W):=h(S=0,W)=\frac{2\sigma}{(1-4W)^{3/2}}
       \left[\tfrac{1}{2}-2W+W^2+\tfrac{1}{2}(1-2W)\sqrt{1-4W}
       \right]^{\delta/2}.
}
Then, we can show that $h^2-h_0^2$ is expressed as
\Eqr{
&h(S,W)^2-h_0(W)^2
  &=\left[16\delta(\delta-1)W^2+\Order{W^3}\right]\sigma^2X^2 \notag\\
&&\qquad    -2(\delta-1)(\delta-2)^2(\delta-3)\sigma^2X^4
    +\Order{WX^4,X^6}
   \notag\\
&&=\delta(\delta-1)W^\delta S^2
   -\frac{(\delta-1)(\delta-2)^2(\delta-3)}{128\sigma^2}W^{2\delta-4}S^4
  \notag\\
&&\qquad  +\Order{W^{\delta+1},W^{2\delta-3}}.
}
From this, it follows that for $\delta=3$ or $\delta\ge4$, $h^2$ is expressed in terms of a function $k(S,W)$ that is continuous at $W=0$ as
\Eq{
h^2(S,W)=h^2_0(W)+W^\delta k(S,W).
\label{k:def}}
Next, for $\delta=2$, $h^2$ can be expanded with respect to $W$ as
\Eq{
h^2
=4\sigma^2\left[1+4W+\left(20-\frac{S^2}{2\sigma^2}
    +\frac{S^4}{64\sigma^4}\right)W^2
    +\Order{W^3}\right].
}
Hence, the expression \eqref{k:def} also holds in this case with a 
function $k$ that is smooth at $W=0$.

%T3>fig:MetricSignature
\begin{figure}[t]
\includegraphics*[height=6cm]{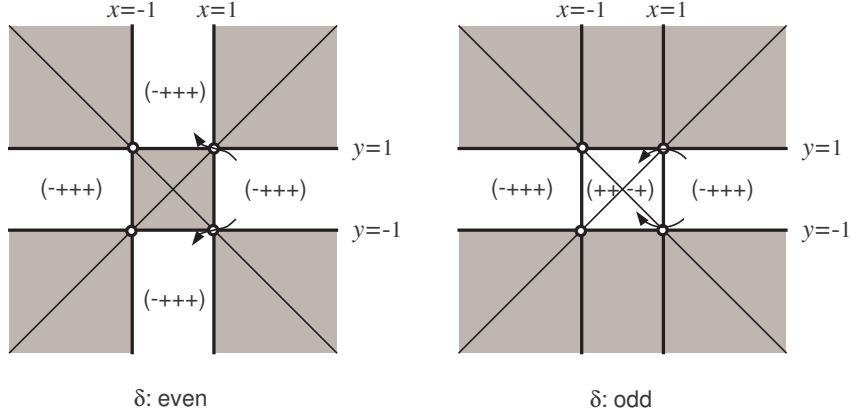}\caption
{\label{fig:MetricSignature} The signature of the metric in the 
$x-y$ plane for the case $\delta$ is an integer, and the path of the 
spacetime extension across P$_\pm$: $(x,y)=(1,\pm1)$. In the shaded 
regions, the metric has the wrong signature $(+---)$ or its 
permutation.}\end{figure} 

Now, we can construct a regular extension across P$_\pm$ by 
introducing the retarded/advanced-type coordinate 
\Eq{
u_\pm:=t \pm \int^W h_0(W)\frac{dW}{W^\delta}.
}
In term of the coordinate system $(t,\phi,u_\pm,S)$, the ZVW metric is written
\Eq{
ds^2=-aW^\delta du_\pm^2 \pm 2ah_0 dWdu_\pm +akdW^2
     +bS^2d\phi^2+cdS^2.
\label{ZVW:metric:uWS}
}
It is clear from the comment above on the behavior of $a$, $b$, $c$ 
and $k$ that this metric can be regularly extended to the region 
$W<0$ for $\delta=2,3$ and $\delta\ge4$. Here, it is understood that 
$W^\delta$ is replaced by $|W|^\delta$ for non-integer $\delta$, as 
mentioned before. This extension connects the regular solutions in 
two regions with $x>1$ and $-1<x<1$ as shown in Fig. 
\ref{fig:MetricSignature}, and is analytic and unique in the case 
$\delta$ is an integer (the extension path in the $x-y$ plane for a 
non-integer $\delta$ is the same as that in the case $\delta$ is an 
even integer). In contrast, for non-integer $\delta$ such that 
$\delta>4$, the extension is at least $C^{[\delta]-2}$, where 
$[\delta]$ represents the maximum integer less than or equal to 
$\delta$, but the $([\delta]-1)$-th derivatives of the metric are 
divergent at $W=0$ in general. 

\begin{figure}[t]
  \includegraphics*[width=\minitwocolumn]{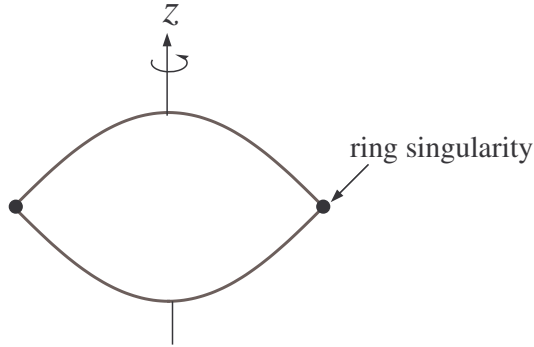}
  \caption{The shape of a spatial section of the horizon for the 
  $\delta=2$ ZVW spacetime, conformally embedded in the Minkowski 
       spacetime $E^{2,1}$.  The ring singularity marked by a full  
       circle actually has an infinite circumference.}
      \label{fig:W2horizon}
\end{figure}

Thus, we have shown that the ZVW spacetime with $\delta=2,3$ or 
$\delta\ge4$ has Killing horizons at P$_\pm$:  
$(\rho,z)=(0,\pm\sigma)$. These two horizons share the ring 
singularity corresponding to the open segment I as the common 
boundary. As a whole, they can be regarded as a single spherical 
horizon with a ring singularity at the equator whose circumference 
is infinite, as is illustrated in Fig. \ref{fig:W2horizon}. Since 
the $W$-derivative of $g_{tt}$ vanishes at $W=0$, this horizon is 
degenerate in its regular part in the sense that the surface gravity 
vanishes. 

Here, note that the four-dimensional 
extension constructed here coincides with the two-dimensional 
extension along the $z$-axis constructed by Papadopoulos et 
al\cite{Papadopoulos.D&Stewart&Witten1981}. In particular, the 
conformal diagram of this two-dimensional sector of the extended 
spacetime is given by Fig. 3 for even $\delta$ and by Fig. 4 for odd 
$\delta$ in their paper. We can also easily draw a conformal diagram 
of the two-dimensional sector corresponding to the 
equatorial plane. The result is just the same as the diagram for a 
negative-mass Schwarzschild black hole, for any $\delta\not=1$. 
Hence, the singularity at the segment I is time-like.

Finally, we comment on the area of the horizon. From \eqref{ZVW:metric:uWS}, its area $A$ for $\delta>2$ is infinite:
\Eq{
A=\lim_{W\tend0}2\times 2\pi \int_0^\infty \sqrt{b}SdS=+\infty.
}
In contrast, for $\delta=2$, it is given by the finite value
\Eq{
A=\lim_{W\tend0}2\times 2\pi\int_0^{4\sigma}\sqrt{bc}SdS
 =32\pi\sigma^2.
\label{ZVW:HorizonArea}
}
This horizon area for the $\delta=2$ case is half of that for the 
Schwarzschild black hole with the same mass $M=2\sigma$ and equal to 
twice of that for the Schwarzschild black hole with mass $M/2$. To 
be precise, we can show that the Schwarzschild solution with mass 
$M=2\sigma$ can be represented by the Israel-Kahn 
solution\cite{Israel.W&Kahn1964} with two black holes of the same 
mass whose centers are separated by $2\sigma$ in the Weyl 
coordinates, and when these centers come closer, the total horizon 
area decreases monotonically. The above area of the $\delta=2$ ZVW 
solution is obtained in the limit the two centers coincide.

%T3>fig:W2horizon

%T1>TS spacetime
%%-------------------------------------------
\section{Tomimatsu-Sato spacetime}
%%-------------------------------------------

In this section, we extend the analysis on the ZVW solution to the 
Tomimatsu-Sato solution. In the present paper, we only consider TS2, 
i.e., the $\delta=2$ case.

The metric of TS2 is expressed in the prolate spheroidal coordinates 
as\cite{Tomimatsu.A&Sato1972}
\Eqr{
&ds^2
 &=-fdt^2+2f\omega dtd\phi+R^2d\phi^2
     +\Sigma^2\left(\frac{dx^2}{x^2-1}+\frac{dy^2}{1-y^2}\right),
\label{TS2:metric:xy}
}
with
\Eq{
f=\frac{A}{B},\quad 
f\omega=\frac{4\sigma q}{p}\frac{(1-y^2)C}{B},\quad \Sigma^2=\frac{B}{p^{4}(x^2-y^2)^3},\quad
R^2=\frac{\sigma^2}{p^2}\frac{(1-y^2)D}{B}.
}
Here, $A,B$, $C$ and $D$ are polynomials given by
\Eqrsub{
&A= &(1-y^2)^4g(Z)g(-Z), 
\label{TS2:A}\\
&B= &(g(x)+q^2y^4)^2+4q^2y^2\left[px^3+1-(px+1)y^2\right]^2,
\label{TS2:B}\\
&C= &q^2(px+1)y^4(-y^2+3)+\left\{-2q^2(px^3+1)+(px+1)g(x)\right\}y^2
  \notag\\
&& -(2px^3-px+1)g(x),
\label{TS2:C}\\
&D=& p^2q^2(x^2-1)y^8+4q^2(px+1)(p^3x^3+3p^2x^2-p^3x+4px-3p^2+4)y^6
 \notag\\
&& -2q^2(3p^4x^6+4p^3x^5-3p^4x^4+8p^3x^3+37p^2x^2
    -12p^3x+48px-13p^2+24)y^4
 \notag\\
&& +4q^2(p^4x^8-p^4x^6+4p^3x^5-4p^3x^3+15p^2x^2+24px-3p^2+12)y^2
\notag\\
&&+g(x)(p^4x^6+6p^3x^5-p^4x^4+16p^2x^4-12p^3x^3+32px^3 \notag\\
&&\quad  +15p^2x^2+6p^3x-15p^2+16),
\label{TS2:D}
}
where
\Eq{
g(x)=p^2x^4+2px^3-2px-1,\quad
Z^2:=\frac{x^2-y^2}{1-y^2}.
\label{g:def}
}

This solution is asymptotically flat, and the parameters $\sigma,p$ 
and $q$ are related to its mass $M$ and angular momentum $J$ as 
\Eq{
 p^2+q^2=1,\quad M=\frac{2}{p}\sigma,\quad J=M^2q.
}
In particular, in the non-rotating case $q=0$, this solution reduces 
to the $\delta=2$ ZVW solution, which is referred to as ZVW2 from 
this point. In the present paper, we only consider the case $0<p, 
q<1$.

%T3>fig:TS2:structure
\begin{figure}[t]
  \includegraphics*[height=\minitwocolumn]{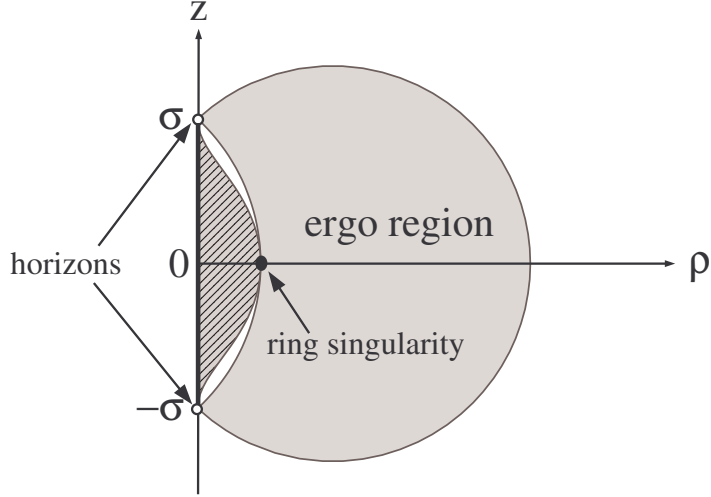}
  \caption{The structure of the Tomimatsu-Sato spacetime with   
  $\delta=2$ and $p=1/\sqrt{2}$. In the hatched gray region, the 
  rotational Killing vector $\partial_\phi$ becomes time-like and 
  its orbits give  closed timelike curves.}
	 \label{fig:TS2:structure}
\end{figure}

%T2>Basic properties
\subsection{Basic properties}

Although TS2 reduces to ZVW2 in the limit $q=0$, the former solution 
has quite different features compared with the latter%
\cite{Tomimatsu.A&Sato1973,Tomimatsu.A&Sato1973a,%
Gibbons.G&Russell-Clark1973}. 

%T3>Ergo region
\subsubsection{Ergo region}

$f=-g_{tt}=-\xi\cdot\xi$ is not positive definite in the 
asymptotically flat region $x>1, -1\le y\le1$, and the so-called 
ergo region appears%
\cite{Gibbons.G&Russell-Clark1973}. From \eqref{TS2:A} and 
\eqref{TS2:B}, $f$ becomes negative in the region $A<0$. Since 
$g(x)=0$ has two single real roots $-x_1$ and $x_0$ with $1<x_0<x_1$ 
for $0<p<1$, the ergo region in $x\ge1$ is given by%
\Eq{
x_0^2-(x_0^2-1)y^2<x^2<x_1^2-(x_1^2-1)y^2.
}
Note that since $Z$ in \eqref{g:def} is expressed in terms of the 
$(X,Y)$ coordinates in \eqref{XYcoord:def} as $Z^2=1+1/X^2$, the 
boundaries of this ergo region, i.e., the infinite redshift 
surfaces, are represented as $X=$const (see Fig. 
\ref{fig:TS2:structure}). 

%T3>Causality violation
\subsubsection{Causality violation}

Since $D$ is not positive definite in $x\ge1$, the norm $R^2$ of the 
rotation Killing vector $\eta=\partial_\phi$ becomes negative in 
some region\cite{Gibbons.G&Russell-Clark1973}. Each $S^1$ orbit in 
this region becomes a closed time-like curve. This region contains 
the segment I: $\rho=0, -\sigma<z<\sigma$, where $\rho$ and $z$ are 
the Weyl-type coordinates related to $x$ and $y$ by 
\eqref{eq:prolate}, and contacts with the inner boundary of the ergo 
region from \eqref{TS2:D}, as shown in Fig. \ref{fig:TS2:structure}. 
 The appearance of this causality violating region in the domain of 
outer communication is the most exotic feature of TS2. 

%T3>Singularity
\subsubsection{Singularity}\label{sec:TS2:singularity}

From \eqref{TS2:metric:xy}, TS2 may have curvature singularity at 
points where $B=0$, in addition to the points $x=\pm1$ and $x=\pm y$ 
in the Weyl case. This is confirmed by looking at the behavior of 
the Weyl curvature. As in the case of the ZVW solutions, 
$\Psi_1=\Psi_3=0$ with respect to the null complex tetrad 
\eqref{ZVW:NullTetrad} with $dt$ replaced by $ dt-\omega d\phi$, 
and all polynomial curvature invariants are expressed as polynomials 
of $\Psi_2$ and $\Psi_0\Psi_4$. By symbolic computations, we can 
show that they have the forms
\Eq{
\Psi_2=\frac{P_1}{B^3},\quad
\Psi_0\Psi_4=\frac{P_2}{B^6},
}
where $P_1$ and $P_2$ are polynomials of $x$ and $y$ that have no 
non-trivial common divisor with $B$. Since the explicit expressions 
for these curvature invariants are quite long, we only give their 
values at the equatorial plane $y=0$:
\Eqrsub{
&& \Psi_2(x,y=0) = \frac{p^4x^6(-2-3px+px^3)}{g(x)^3},\\
&& (\Psi_0\Psi_4)(x,y=0)=\frac{9p^8x^{10}
   (x^8p^2-2p^2x^6+4x^5p+5p^2x^4-4px^3+8px+4)}{g(x)^6}.
}
From these expressions, it follows that the curvature invariants 
diverge at the zero of $g(x)$, $x=x_0$, which is just the 
intersection of the inner boundary of the ergo region with the 
equatorial plane. It is easy to see from \eqref{TS2:B} that $B$ 
vanishes only at $(x,y)=(x_0,0)$ for $x\ge1$ and $0<p<1$, except for 
the two points P$_\pm$: $(x,y)=(1,\pm1)$. Hence, in contrast to the 
ZVW spacetime, there exists no curvature singularity on the open 
segment I, and instead there appears the famous ring singularity%
\cite{Tomimatsu.A&Sato1972}, which has an infinite circumference, 
because $x=x_0$ is a single root of $D$ and a double root of $B$ on 
the equatorial plane $y=0$ from \eqref{TS2:B} and \eqref{TS2:D}. 
This ring singularity reduces to the ring-like singularity at the 
segment I of ZVW2 in the limit $q\tend0$.
 
$B$ also vanishes at P$_\pm$. However, we can show by symbolic 
computations that the curvature invariants have finite limits at 
these points. These limits are direction dependent as in the case of 
the ZVW spacetime. For example, their values in the cases in which 
one approaches P$_\pm$ from $|z|>\sigma$ and from $|z|<\sigma$ along 
the $z$-axis are given by 
\Eqrsub{
&X=0,Y=\pm1:& \Psi_2=-\frac{p(p\pm iq)}{8(1+p)},\quad
              \Psi_0\Psi_4=\frac{9p^2(p\mp iq)^2}{64(p+1)^2},\\
&X=\infty,Y=\pm1: & \Psi_2=\frac{p^4(-q\pm ip)}{8q^3(1+p)},\quad
                 \Psi_0\Psi_4=-\frac{9p^8(p\mp iq)^2}{64q^6(p+1)^2}.
}
Thus, P$_\pm$ are quasi-regular 
singularities\cite{Gibbons.G&Russell-Clark1973,Ernst.F1976}. In the 
next subsection, we will show that these two points are disconnected 
Killing horizons, by constructing regular extensions of the 
spacetime across these points.

%T4>fig:tildet=const
\begin{figure}[t]
\includegraphics*[height=6cm]{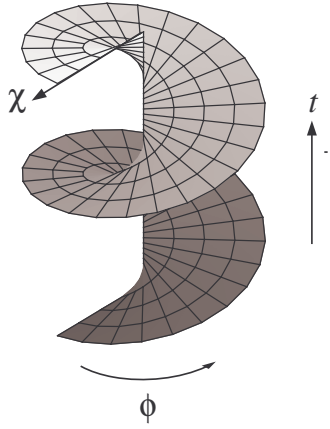}
\caption{$\tilde t=$const surfaces}
\label{fig:tildet=const}
\end{figure}

All metric components of \eqref{TS2:metric:xy} as well as the 
polynomial curvature invariants are regular and finite on the open 
segment I, which is expressed as $x=1,-1<y<1$ in the $(x,y)$ 
coordinates, in contrast to the case of ZVW2. However, the spacetime 
turns out to be singular on this 
segment\cite{Gibbons.G&Russell-Clark1973}. To see this, let us 
rewrite the metric \eqref{TS2:metric:xy} as 
\Eq{
ds^2=-f(dt-\omega d\phi)^2
    + \Sigma^2\left(d\chi^2+\ell^2\sinh^2\chi d\phi^2\right)
    +\Sigma^2\frac{dy^2}{1-y^2},
\label{TS2:metric:chiy}
}
where $\chi$ is related to $x$ by $x=\cosh\chi$, and $\ell$ is defined by
\Eq{
\ell^2=\frac{\sigma^2(1-y^2)}{f\Sigma^2}.
}
In the neighbourhood of the segment I, $\omega$ behaves as
\Eqr{
&& \omega=\frac{4\sigma(p+1)}{qp}+\omega_1(x,y)\sinh^2\chi;\notag\\
&& \omega_1=-\frac{4\sigma (p+1)[(p-1)(y^4+1)+2(3p+1)y^2]}{q^3(1-y^2)^2}
       +\Order{\chi^2}.
}
Hence, if we introduce the new time coordinate $\tilde t$ by
\Eq{
\tilde t=t - \omega_0\phi;\quad \omega_0=\frac{4\sigma(p+1)}{qp},
}
\eqref{TS2:metric:chiy} can be written as
\Eq{
ds^2=-f(d\tilde t-\omega_1\sinh^2\chi d\phi)^2
    + \Sigma^2\left(d\chi^2+\ell^2\sinh^2\chi d\phi^2\right)
    +\Sigma^2\frac{dy^2}{1-y^2}.
}
From this, it follows that the spacetime is regular on the segment 
I, if $\ell$ is unity on I. However, since $\ell$ behaves around I 
as 
\Eq{
\ell=\frac{p^2}{q^2}+\frac{p^2(p^2+3)}{q^4(1-y^2)}\sinh^2\chi
     +\Order{\chi^4},
}
the spacetime is not regular on I unless $p=q=1/\sqrt{2}$ and has a 
conical singularity there. The deficit angle is positive for 
$q^2>p^2$ and negative for $q^2<p^2$.

This is not the whole story. There is another subtlety concerning 
the spacetime structure around I\cite{Gibbons.G&Russell-Clark1973}. 
Since $\phi$ is an angle coordinate and defined modulo $2\pi$, the 
new time coordinate $\tilde t$ is not a single-valued function, and 
each $\tilde t=$const surface has a helical structure in the 
original coordinates, as illustrated in Fig. \ref{fig:tildet=const}. 
These helical surfaces cannot be extended regularly to the axis 
$\chi=0$. Hence, $\phi$ cannot be regarded as an angle coordinate on 
these $\tilde t=$const surfaces, but rather should be regarded as a 
parameter of helical orbits of the Killing vector 
$\partial_\phi+\omega_0\partial_t$. In this sense, the deficit angle 
interpretation above does not have a global meaning. This exotic 
structure arises because the two Killing vectors $\partial_t$ and 
$\partial_\phi$ become parallel on I.

%T3>Komar mass
\subsubsection{Komar mass}

%T4>fig:TS2:My
\begin{figure}[t]
\includegraphics*[height=6cm]{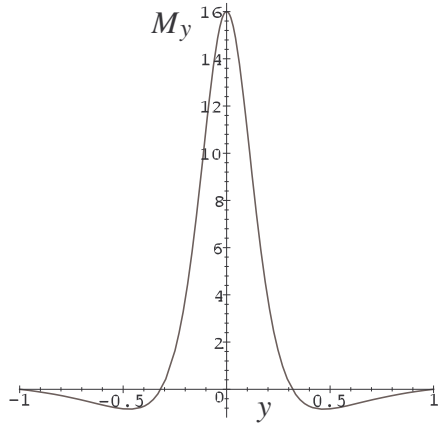}
\caption{The behavior of $M_y$ at $x=1$ for $p=0.5$}
\label{fig:TS2:My}
\end{figure}

For the metric \eqref{TS2:metric:xy}, the Komar mass 
\eqref{KomarMass:def} contained in a 3-surface that is bounded by 
a closed 2-surface determined by a curve $\gamma$ in the $x-y$ plane 
is expressed as
\Eq{
M= \int_\gamma \left(M_x dx+M_y dy\right),
\label{KomarMass:TS:general}
}
where
\Eq{
(1-y^2)M_y dx - (x^2-1)M_x dy
=\frac{1}{4\sigma}\left[R^2 df + f\omega d(f\omega)\right].
}
We can show that for any curve $\gamma$ whose end points are on the 
regular part of the $z$-axis, $z>\sigma$ and $z<-\sigma$, $M$ is 
equal to the total mass $M=2\sigma/p$, if $\gamma$ does not pass the 
ring singularity. This implies that the ring singularity has zero 
gravitational mass. In contrast, from the asymptotic behavior of 
$M_x$ and $M_y$ around $x=1$, 
\Eqrsub{
&& M_x=\frac{4\sigma y[(1-p)(y^6-y^4)+(p+7)y^2+9-p]}
      {[(1-p)y^4+2(p+3)y^2+1-p]^2} +\Order{x-1},\\
&& M_y=\frac{4\sigma (1-y^2)[(1-p)(y^4+1)-2(3p+1)y^2]}
     {p[(1-p)y^4+2(p+3)y^2+1-p]^2} + \Order{x-1},
}
the main contribution to the integral on the right-hand side of 
\eqref{KomarMass:TS:general} comes from the segment I for a curve 
close to I (see Fig. \ref{fig:TS2:My}). This seems to indicate that 
the gravitational mass is carried by the singularity on I. However, 
this interpretation may not be valid, because a 2-surface 
corresponding to a curve close to the segment I is time-like. 
Furthermore, $M_y$ is not positive definite, as shown in Fig. 
\ref{fig:TS2:My}.

%T2>Extension across P+-
\subsection{Extension across P$_\pm$}

In order to construct an extension of TS2, we utilize the 
coordinates $(X,W)$, where $X$ is the coordinate defined in 
\eqref{XYcoord:def} and $W$ is defined in terms of the $Y$ 
coordinate there by 
\Eq{
W=\frac{1}{4}(1-Y^2).
}
Then, we can show that the metric is written in terms of functions 
$a(X,W)$, $b(X,W)$, $c(X,W)$, $h(X,W)$ and $\Omega(X,W)$ that are 
regular analytic in a neighbourhood of $W=0$i$Y=\pm1$) as 
\Eq{
ds^2=-aW^2\left(dt^2-h^2\frac{dW^2}{W^4}\right)
     +b X^2(d\phi-\Omega dt)^2+c dX^2.
}
The asymptotic behavior of these functions at $W=0$ are given by
\Eqrsubl{TS2:metric:P}{
&& a=\frac{2(p^2+q^2X^4)}{(p+1)(1+X^2)^2} 
     +\frac{8(p^2-q^2X^6)}{(p+1)(1+X^2)^3}W +\Order{W^2},\\
&& b=\frac{8\sigma^2(p+1)}{p^2+q^2X^4}+\Order{W},\\
&& c=\frac{8\sigma^2(p+1)(p^2+q^2X^4)}{p^4(1+X^2)^4}+\Order{W},\\
&& h^2=\frac{(p+1)^2}{p^4}\left\{1+4W+4(p+4)W^2\right\} \notag\\
&&\qquad    -\frac{4(p+1)^2
    \left\{2p^3+p^2(2p-1)X^2+q^2X^4\right\}X^2}
     {p^4(1+X^2)^2}W^2+\Order{W^3},\\
&&\sigma\Omega=\frac{pq}{p+1}W
    -\frac{q(p^2+q^2X^4)X^2}{p(p+1)(1+X^2)^2}W^2
    +\Order{W^3}.
}
From this behavior, if we put
\Eq{
h_0(W)=h(X=0,W),\quad
\Omega_0(W)=\Omega(X=0,W), 
}
$h^2$ and $\Omega$ can be expressed in terms of functions $k(X,W)$ 
and  $\Omega_1(X,W)$ that are regular analytic at $W=0$ as 
\Eqrsub{
&& h^2=h_0(W)^2+k(X,W)W^2,\\
&& \Omega=\Omega_0(W)+ \Omega_1(X,W)W^2.
}
Hence, in the advanced/retarded coordinates
\Eqrsub{
&& u_\pm:=t \pm \int^W h_0(W)\frac{dW}{W^2},\\
&& \phi_\pm:=\phi \pm \int^W \Omega_0(W)h_0(W)\frac{dW}{W^2}, 
}
the metric takes the form that is regular analytic at $W=0$:
\Eq{
ds^2=-aW^2du_\pm^2\pm 2h_0 du_\pm dW+ akdW^2
     +bX^2(d\phi_\pm\pm \Omega_1h_0dW-\Omega du_\pm)^2
     +cdX^2.
\label{TS2:extension}
}
Since $W=0$ is a null hypersurface and the Killing vector 
$\xi=\partial_t$ is its null tangent, the hypersurface $W=0$ is a 
Killing horizon. Because $W=0$ is the double root of $\rho^2/R^2$, 
the horizon is degenerate, i.e., the surface gravity vanishes. 
Further, since $\Omega$ vanishes at $W=0$, the horizons are 
non-rotating.

%T2>Area and shape of the horizons
\subsection{Area and shape of the horizons}

It is easy to estimate the area of the horizons. From 
\eqref{TS2:metric:P} and \eqref{TS2:extension}, we obtain 
\Eq{
A=2\times \lim_{W\tend0} 2\pi\int_0^\infty \sqrt{bc}XdX
 =\frac{16\pi \sigma^2 (p+1)}{p^2},
\label{TS2:HorizonArea}
}
which coincides with \eqref{ZVW:HorizonArea} for the non-rotating 
limit $p=1$. Since it can be written as
\Eq{
A=4\pi M(M+\sqrt{M^2-a^2})
}
with $a=J/M$, it is half of the area of the Kerr black hole with the 
same mass $M$.

\begin{figure}[t]
\includegraphics*[height=6cm]{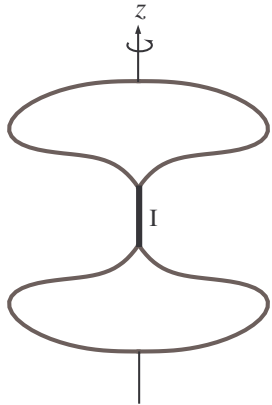}
  \caption{The shape of a spatial section of the horizon for TS2 
  with $p=0.6$.}\label{fig:TS2:horizon}
\end{figure}

Although the horizon area was finite in spite of its non-compactness 
in the case of ZVW2, the horizons become compact in the case of TS2. 
In order to see this, let us rewrite the metric of the 
two-dimensional section $W=0, u_\pm=$const of the horizon as 
\Eq{
ds_H^2=\frac{2\sigma^2(p+1)}{p^2\ell(\theta)}
 \left[d\theta^2+\ell^2(\theta)\sin^2\theta d\phi_\pm^2\right],
}
where
\Eqr{
&& X=\tan\frac{\theta}{2} \ (0\le \theta \le\pi),\\
&& \ell(\theta)=\frac{4p^2}{(\cos\theta+p^2-q^2)^2+4p^2q^2}.
}
For $0<p,q<1$ with $p^2+q^2=1$, $\ell(\theta)$ is positive and 
finite for $0\le\theta\le\pi$. Hence, each horizon at P$_\pm$ is 
compact and homeomorphic to $S^2$. Further, since $\ell(0)=1$, it is 
smooth for $0\le\theta<\pi$. However, since $\ell(\pi)=p^2/q^2$, it 
has a conic singularity at $\theta=\pi$, i.e., at the point where it 
contacts with the segment I, except for the case $p=q=1/\sqrt{2}$, 
as illustrated in Fig. \ref{fig:TS2:horizon}. The deficit angle is 
the same as that at the open segment I discussed in 
\ref{sec:TS2:singularity}; it is positive for $q^2>p^2$ and negative 
for $q^2<p^2$. In particular, in the limit $q\tend0$, the deficit 
angle tends negative infinity, which is consistent with the fact 
that each regular portion of the horizon is a half sphere for ZVW2.

%T1>Summary and Discussions
%%-------------------------------------------
\section{Summary and Discussions}
%%-------------------------------------------

We have investigated the structure of the domains of the outer 
communication and their boundaries for the Zipoy-Voorhees-Weyl 
solution with arbitrary $\delta$ and for the Tomimatsu-Sato solution 
with $\delta=2$ and $0<p,q<1$(TS2). We have shown that three types 
of singularities arise for these spacetimes. One is the naked 
curvature singularity. It appears on the segment I: $\rho=0,
 -\sigma<z<\sigma$ for the ZVW spacetime with $\delta\not=0,1$. We 
have shown that this singularity is point-like for $\delta<0$, 
string-like for $0<\delta<1$ and ring-like for $1<\delta$. The last 
feature has not been noticed so far, and has a close relation with 
the existence of the ring singularity for TS2. Another is an 
inextendible 
quasi-regular singularity at the segment I for TS2. We have 
confirmed that this singularity is conical locally, but has a more 
complicated helical structure globally, as was pointed out by 
Gibbons and Russell-Clark\cite{Gibbons.G&Russell-Clark1973}. The 
third is the directional quasi-regular singularities at the two 
points P$_\pm$: 
$(\rho,z)=(0,\pm\sigma)$ for the ZVW spacetime with $\delta\ge2$ 
and for TS2. We have proved that these points are degenerate Killing 
horizons for the ZVW spacetime with $\delta=2,3$ or $\delta\ge4$ 
and for TS2, by constructing regular extensions across these points 
explicitly. 

An  interesting feature of the horizon for the ZVW spacetime is that 
the two regular horizons at P$_\pm$ share the ring singularity at I 
as the common boundary and these pieces altogether form a single 
spherical horizon with a ring singularity at the equator. In 
contrast, although the two horizons of TS2 also contact with the 
conical segment I, each of them is a compact surface which is 
homeomorphic to $S^2$ and has a conical singularity at the point 
where it contacts I for $p\not=q$. Thus, it is natural to regard 
that the horizon of TS2 has two disconnected components. This result 
is rather surprising because TS2 can be obtained from the 
Kramer-Neugebauer solution representing a superposition of two Kerr 
solutions\cite{Kramer.D&Neugebauer1980}, as the limit when the 
centers of two black holes coincide \cite{Oohara.K&Sato1981}. This 
may indicates a new possibility for the final states of 
gravitational collapse, if the cosmic censorship does not hold.

In this connection, it should be noted that the 
horizon of the ZVW solution with $\delta=2$(ZVW2)  practically has a 
single component, as pointed above, although this solution can be 
also obtained as a similar limit from the Israel-Khan 
solution\cite{Israel.W&Kahn1964} with two black holes with the same 
mass. Further, we have pointed out that the naked curvature 
singularity of the ZVW spacetime can be regarded to have a positive 
gravitational mass. This should be contrasted with the fact that the 
ring singularity of TS2 has zero mass.

%T1>Acknowledgement
\section*{Acknowledgement}

H.K. would like to thank Misao Sasaki for valuable comments. This 
work was partly supported by the Grand-In-Aid for scientific 
researches by JSPS (Number. 11640273). 

%T1>Appendix
\appendix
\section{Geodesic equation in the ZVW spacetime}

Geodesics in the ZVW spacetime are determined as solutions to the 
following set of equations: 
\Eqrsubl{ZVW:GeodesicsEq}{
&& \dot t=E\pfrac{x+1}{x-1}^\delta,\\
&& \dot\phi=L\frac{(x-1)^{\delta-1}}{(x+1)^{\delta+1}(1-y^2)},\\
&& \frac{\dot x^2}{x^2-1}+\frac{\dot y^2}{1-y^2}
   = \frac{(x^2-y^2)^{\delta^2-1}}{(x^2-1)^{\delta^2}} F,
\label{ZVW:GeodesicEq:affine}\\
&& F \dfrac{d^2y}{dx^2}
  = F\frac{(\delta^2-1)y(1-y^2)}{(x^2-1)(x^2-y^2)}
    - L^2\frac{y(x-1)^{2\delta-2}}{(1-y^2)(x+1)^{2\delta+2}}
 \notag\\
&& \qquad +\left[\epsilon\frac{(x-1)^{\delta-1}}{(x+1)^{\delta+1}}
     -F \frac{2(x-\delta)(x^2-y^2)+(\delta^2-1)x(1-y^2)}
        {(x^2-1)(x^2-y^2)}
     \right.\notag\\
&&\left.\qquad
   -L^2\frac{(x-2\delta)(x-1)^{2\delta-2}}{(1-y^2)(x+1)^{2\delta+2}}
   \right]\frac{dy}{dx}
 \notag\\
&&\qquad  -\left[F\frac{y\{x^2-y^2+(1-\delta^2)(1-y^2)\}}{(1-y^2)(x^2-y^2)}
    +L^2\frac{y(x-1)^{2\delta-1}}{(1-y^2)^2(x+1)^{2\delta+1}}
    \right]\pfrac{dy}{dx}^2
 \notag\\
&&\qquad  +\left[\frac{\delta\epsilon}{1-y^2}\pfrac{x-1}{x+1}^\delta
   -F\frac{x(x^2-1)+\delta^2 x(1-y^2)-2\delta (x^2-y^2)}
     {(1-y^2)(x^2-y^2)} \right.\notag\\
&&\left.\qquad  -L^2\frac{(x-2\delta)(x-1)^{2\delta-1}}
      {(1-y^2)^2(x+1)^{2\delta+1}} \right]\pfrac{dy}{dx}^3,
\label{ZVW:GeodesicEq:orbit}}
where the dot denotes the derivative with respect to an affine 
parameter, and $F$ is given by
\Eq{
F:=E^2 - L^2\frac{(x-1)^{2\delta-1}}{(1-y^2)(x+1)^{2\delta+1}}
      -\epsilon\pfrac{x-1}{x+1}^\delta.
}
$E$ and $L$ are the specific energy and the specific angular 
momentum of a particle, and $\epsilon$ is equal to $1$ for time-like 
geodesics and $0$ for null geodesics.

%T1>Rererences
%\bibliographystyle{jphys_n_usrt}
%\bibliography{WTS2}

\end{document}